\documentclass[a4paper]{jpconf}
\usepackage{graphicx}
\begin{document}
\newcommand{\bef}{\begin{figure}[!htb]}
\newcommand{\eef}{\end{figure}}
\def \lt {\mbox{$\ <\ $}}
\def \gt {\mbox{$\ >\ $}}
\def \la {\langle}
\def \ra {\rangle}
\def \auau  {$Au + Au$ }
\def \cucu  {$Cu + Cu$ }
\def \pp    {$p + p$ }
\def \npart {$N_{part}$ }
\def \xxi       {$\Xi^- + \overline{\Xi}^+$ }
\def \llam      {$\Lambda + \overline{\Lambda}$ }
\def \oom       {$\Omega^- + \overline{\Omega}^+$ }
\def \etal   {\mbox{$\mathrm{\it et\ al.}$}}
\newcommand{\mean}[1]{\left\langle #1 \right\rangle}
\newcommand{\Red}[1]{\textcolor[named]{Red}{#1}}
\newcommand{\Magenta}[1]{\textcolor[named]{Magenta}{#1}}
\newcommand{\Blue}[1]{\textcolor[named]{Blue}{#1}}
\newcommand{\Green}[1]{\textcolor[named]{Green}{#1}}
\widowpenalty=10000
\clubpenalty=10000
\title{Beam Energy Dependence of Directed and Elliptic Flow Measurement from the STAR Experiment}
\author{\bf{Yadav Pandit}  (for the STAR Collaboration) }
\address{ Department of Physics, Kent State University, USA }
\ead{ ypandit@kent.edu}

\begin{abstract}
 Measurements of anisotropic flow in heavy-ion collisions  provide insight into the early stage of the system's evolution. This proceedings presents directed and elliptic flow for Au+Au collisions at 39, 11.5 and 7.7 GeV, and for Cu+Cu at 22.4 GeV, measured in the
 STAR Experiment at RHIC. Differential measurements of directed and elliptic flow of charged particles as a function of centrality, transverse momentum and pseudorapidity are discussed.  
\end{abstract}

\section{Introduction}

The study of collective flow in relativistic nuclear collisions has potential to offer insights
into the equation of state of the produced matter \cite{whitepapers, collectiveflow}. Anisotropic 
flow is characterized by the Fourier coefficients \cite{methods}

\begin{equation}
v_n = \langle \cos n( \phi-\Psi_R ) \rangle   ,
\end{equation}

where $\phi$ denotes the azimuthal angle of an outgoing particle, $\Psi_R$ is the orientation 
of the reaction plane and $n$ denotes the harmonic. The reaction plane is defined by the beam axis 
and the impact parameter \cite{methods}.

Directed flow, $v_1$, is the first harmonic coefficient of the above Fourier expansion of the 
final momentum-space azimuthal anisotropy, and it reflects the collective sidewards motion of 
the particles in the final state.  In regions that are closer to beam rapidity (\emph{y}) than to midrapidity, 
directed flow is imparted very early, at a pre-equilibrium stage of the collision \cite{Sorge, 
Herrmann}, and thus it probes the onset of bulk collective behavior.  Both hydrodynamic and nuclear
transport models \cite{Hydro,Transport} indicate that directed flow is a sensitive signature 
for phenomena related to a possible phase transition, especially in the general region of beam energy 
under investigation here \cite{bes}. In particular, the shape of $v_1(y)$ in the midrapidity region is 
of special interest because it has been argued that directed flow may exhibit flatness 
at midrapidity due to a strong, tilted expansion of the source. Such tilted expansion gives rise
to anti-flow  or a 3rd flow ~\cite{Csernai}  component. The anti-flow is perpendicular to the 
source surface, and is in the opposite direction to the repulsive ``bounce off" motion of nucleons. 
If the tilted expansion is strong enough, it can cancel and reverse the motion in the bounce-off direction
and results in a negative $v_{1}(y)$ slope at midrapidity, potentially producing a wiggle-like structure in
$v_1(y)$.  A wiggle for baryons is a possible signature of a phase transition between hadronic matter and
Quark Gluon Plasma (QGP), although QGP is not the only possible explanation~\cite{Brachmann,Stocker,Csernai}. 
If strong but incomplete baryon stopping is assumed together with strong space-momentum correlations caused by
transverse radial expansion, then a wiggle structure might be explained even in a hadronic system~\cite{Wiggle}.

Elliptic flow, $v_2$, is the second harmonic coefficient of the Fourier expansion above.  The 
initial-state spatial eccentricity of the participant zone drives the process whereby the 
interactions produce an anisotropic distribution of momenta relative to the reaction plane.  The 
elliptic momentum anisotropy saturates quite early in the collision evolution, although a little later 
than when directed flow is imparted \cite{Sorge, Herrmann}.  Elliptic flow can provide information about 
the pressure gradients in a hydrodynamic description, and about the effective degrees of freedom, 
the extent of thermalization, and the equation of state of the matter created at early times.  
Studying the dependence of elliptic flow on system size, number of constituent quarks, and 
transverse momentum/mass, are crucial to the understanding of the properties of the produced matter 
\cite{whitepapers}.

\section{Methods and Analysis}

        In this  proceedings, we report $ v_{1} $  and $v_{2} $ measurements by the STAR experiment from $\sqrt{s_{NN}}$  = 39, 11.5 and 7.7 GeV Au + Au and  $\sqrt{s_{NN}}$ = 22.4 GeV
   Cu + Cu collisions. Data were taken from Run 10 (2010) and Run 5 (2005). The STAR Time Projection Chamber (TPC) \cite{startpc} was used as the main detector for charged particle tracking and second order event plane determination for elliptic flow analysis. The centrality was determined by the number of tracks from the pseudorapidity region $ |\eta|  < 0.5 $. Forward Time Projection Chambers (FTPCs) were also used for charged particle  tracking at forward rapidities. Two  Beam Beam Counters  covering  $3.3  < |\eta|  < 5.0$ were used to reconstruct the first-order  event plane for directed flow analysis. The pseudorapidity gap between BBC and TPC allows us to reduce some of the non-flow effects.   Elliptic flow measurements with the standard event plane method are based on the first-order event plane from the BBC as well as the second-order event  plane based on tracks in the main Time Projection Chamber \cite{startpc} of STAR. We also present elliptic flow based on two- and four-particle correlations . We analyzed minimum bias events with  the primary collision vertex position along the beam direction $(V_z)$ within 30,50 and 70 cm of the center of the detector are selected for this analysis for  Au+ Au collisions at 39, 11.5 and 7.7 GeV respectably and 30 cm for Cu+Cu collisions at 22.4 GeV . In order to reject events which involve interactions  with the beam pipe and also to minimize beam-gas interactions,  the event vertex radius ($\sqrt{V_x^2 + V_y^2}$, where $V_x$ and $V_y$ are the vertex positions along the $x$ and $y$ directions, respectively) is required  to be less than 2 cm. For these  analyses,  tracks which have transverse momenta $ p_t > 0.2 $ GeV $/c$, pass within  3 cm of the primary vertex, have at least 15 space points in the main TPC acceptance  $(|\eta| \lt 1.0)$ or 5 space points in the case of tracks in the FTPC acceptance $2.5 \lt |\eta| \lt 4.0$). Also, the ratio of the number of actual space points to the maximum possible number of space points for that track's trajectory was required to be  greater than 0.52, which prevents split tracks from being counted as two tracks.The differential measurement of directed and elliptic flow  is presented as a function of pseudorapidity ($\eta$),  transverse momentum and centrality.    

\bef
\begin{center}
\includegraphics[scale=0.79]{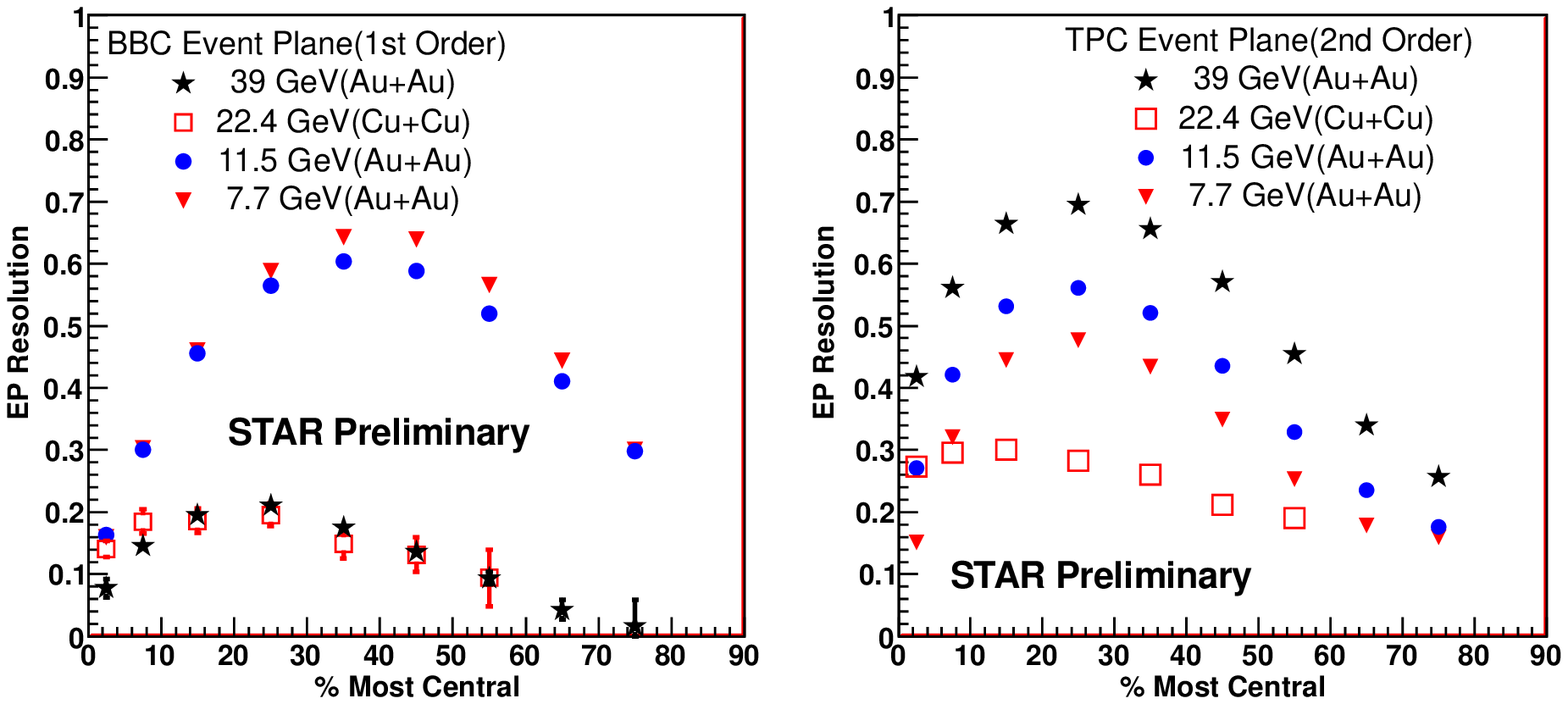}
\caption{ The left panel shows the first-order event plane resolution based on information from the STAR BBC as a function of centrality for Au + Au collisions at $\sqrt{s_{NN}}$ 
= 39, 11.5 and 7.7 GeV and for Cu + Cu collisions at  $\sqrt{s_{NN}}$ = 22.4 GeV. The right panel shows the second-order event plane resolution 
based on the TPC as a function of centrality for Au + Au collisions at $\sqrt{s_{NN}}$ = 39, 11.5 and 7.7 GeV and for Cu + Cu collisions at $\sqrt{s_{NN}} $ = 22.4 GeV. The errors shown are statistical. } 
\label{fig1}
\end{center}
\eef
           
Figure~\ref{fig1}, left panel, shows the first-order event plane resolution as a function of centrality for Au + Au collisions at $\sqrt{s_{NN}}$  = 39, 11.5 and 7.7 GeV and for Cu + Cu collisions at  $\sqrt{s_{NN}} = 22.4$ GeV based on the  event plane reconstructed  using BBC information. The right panel shows the second-order event plane resolution reconstructed using charged tracks from the TPC.  The errors shown are statistical. Generally, the first-order event plane is used for directed flow measurement and the second-order event plane is used for elliptic flow measurement in the standard method of flow analysis. STAR is well suited for these measurements, with very good event plane resolution down to at least 7.7 GeV.

 \section{Results and Discussion}

\subsection{Directed flow}

\bef
\begin{center}
\includegraphics[scale=0.39]{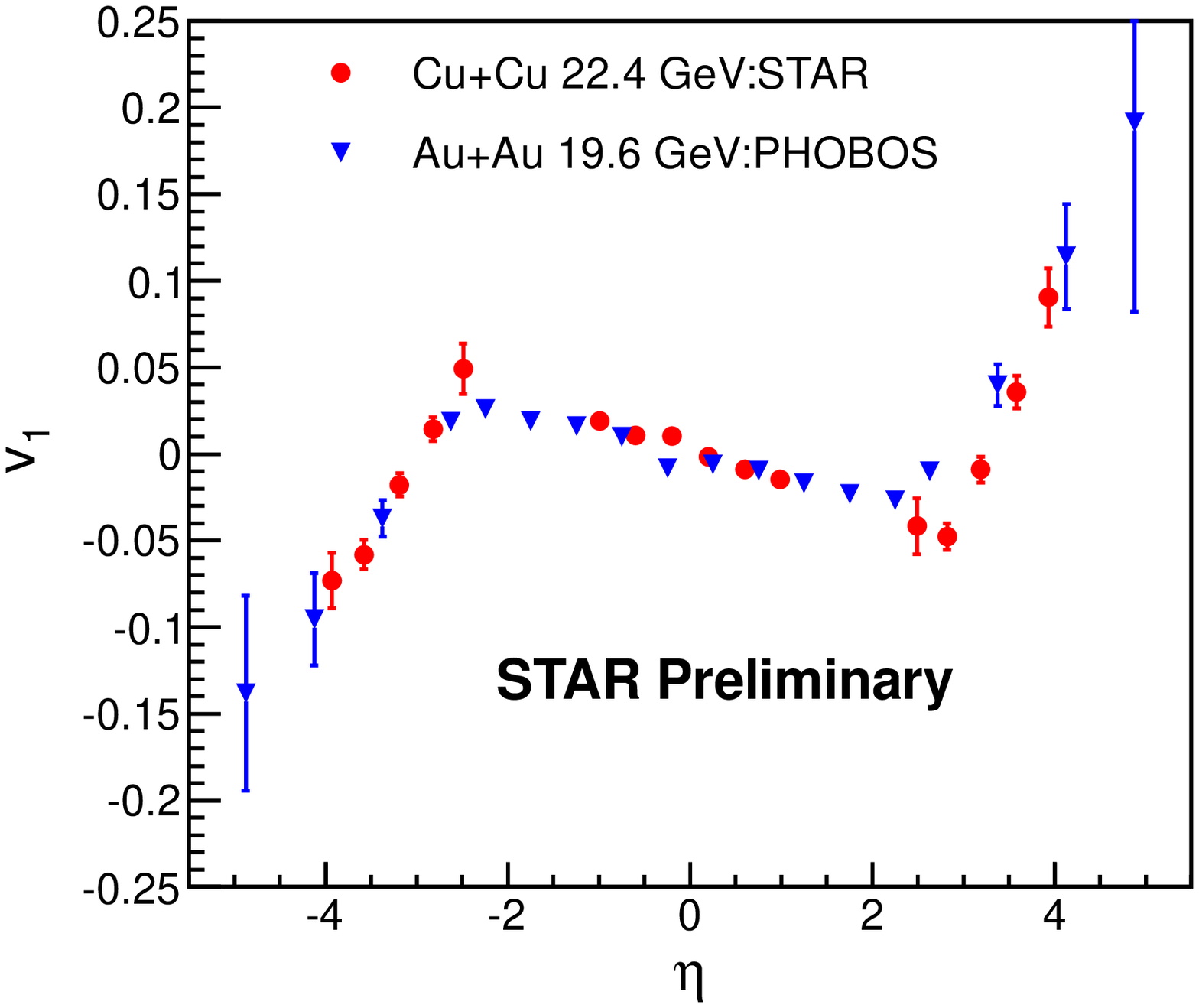}
\caption{Charged hadron $v_1$ vs. $\eta$ for 0--60\% centrality Cu+Cu collisions at $\sqrt{s_{NN}}  = 22.4$ GeV obtained from the standard method based on the BBC event plane. The errors shown are statistical. Results are compared to $v_1$ (solid triangles) from 0--40\% centrality Au+Au collisions at $\sqrt{s_{NN}} = 19.6$ GeV~\cite{PHOBOS_v1} from the PHOBOS collaboration. }
\label{fig2}
\end{center}
\eef
Figure~\ref{fig2} shows charged hadron $v_1\{{\rm BBC}\}$ in Cu+Cu collisions for 0--60\% 
centrality at $\sqrt{s_{NN}} = 22.4$ GeV, compared to 0--40\% centrality Au+Au collisions 
at $\sqrt{s_{NN}} = 19.6$ GeV. It is seen that the latter (PHOBOS) results are quite similar, notwithstanding the difference in system size, and the fact that the centrality range and beam energy are not quite the same.  
At 200 GeV and 62.4 GeV, for 30-60\% central collisions, we have previously reported that directed flow does not vary within 
errors between Au+Au and Cu+Cu   \cite{v1-4systems}.  The new finding reported here suggests that 
this behavior extends to lower energies. 
\bef
\begin{center}
\includegraphics[scale=0.69]{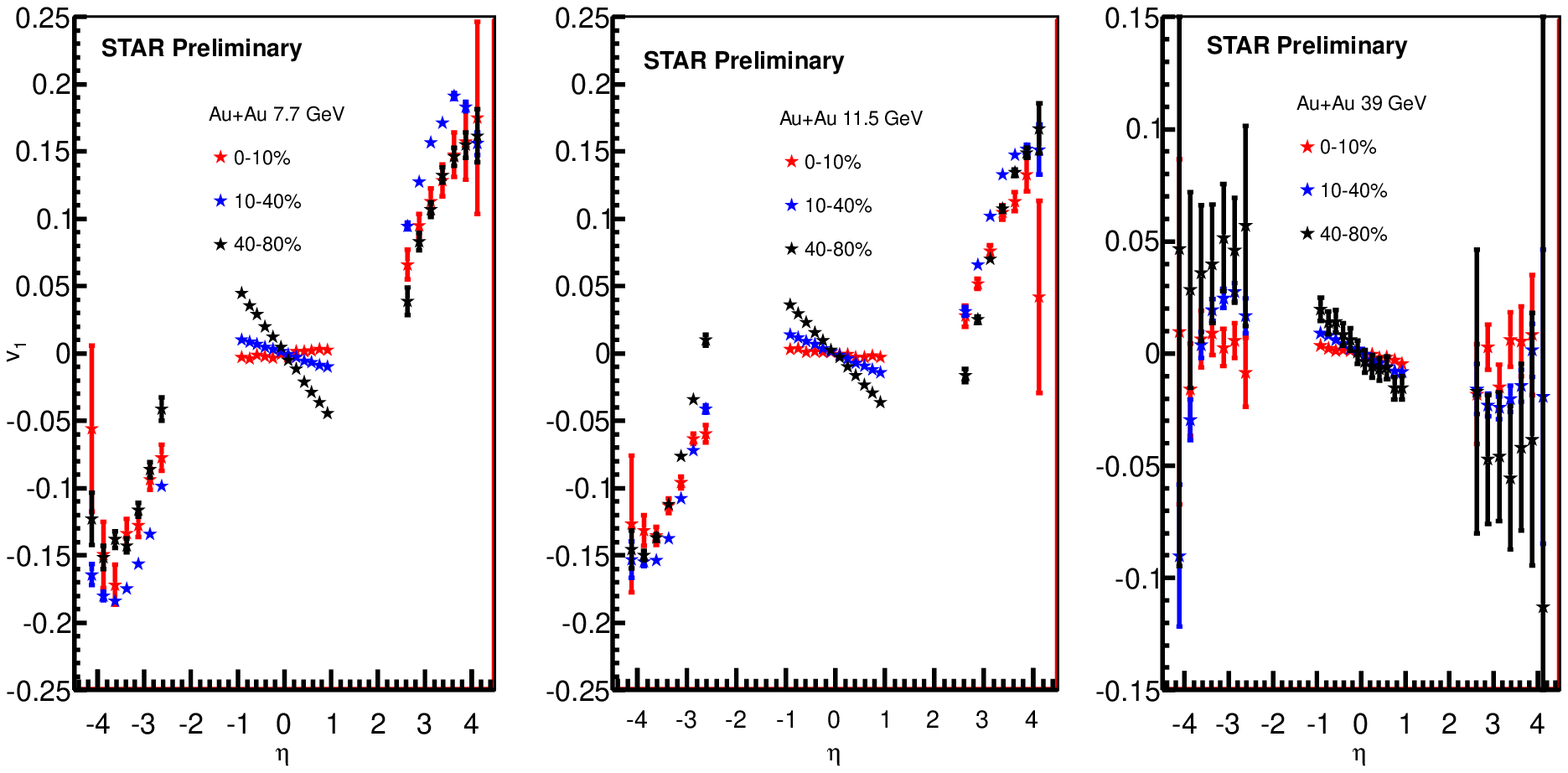}
\caption{Charged hadron $v_{1} $ as a function of pseudorapidity  $\eta$ for central 0--10 \%, mid central 10--40\% and peripheral 40--80 \%  Au+Au collisions  at $\sqrt{s_{NN}}$  = 39, 11.5 and 7.7   GeV. The errors shown are statistical.}
\label{fig3}
\end{center}
\eef
Figure ~\ref{fig3} shows charged hadron $v_{1} $ as a function of pseudorapidity  $\eta$ for central 0--10 \%, mid central 10--40\% and peripheral 40--80 \%  Au+Au collisions  at $\sqrt{s_{NN}}$  = 39, 11.5 and 7.7 GeV. The errors shown are statistical. We observe centrality dependence of the directed flow signal specially prominent at mid rapidity( $ | \eta | <1.0)$  than at forward rapidities.
\bef
\begin{center}
\includegraphics[scale=0.69]{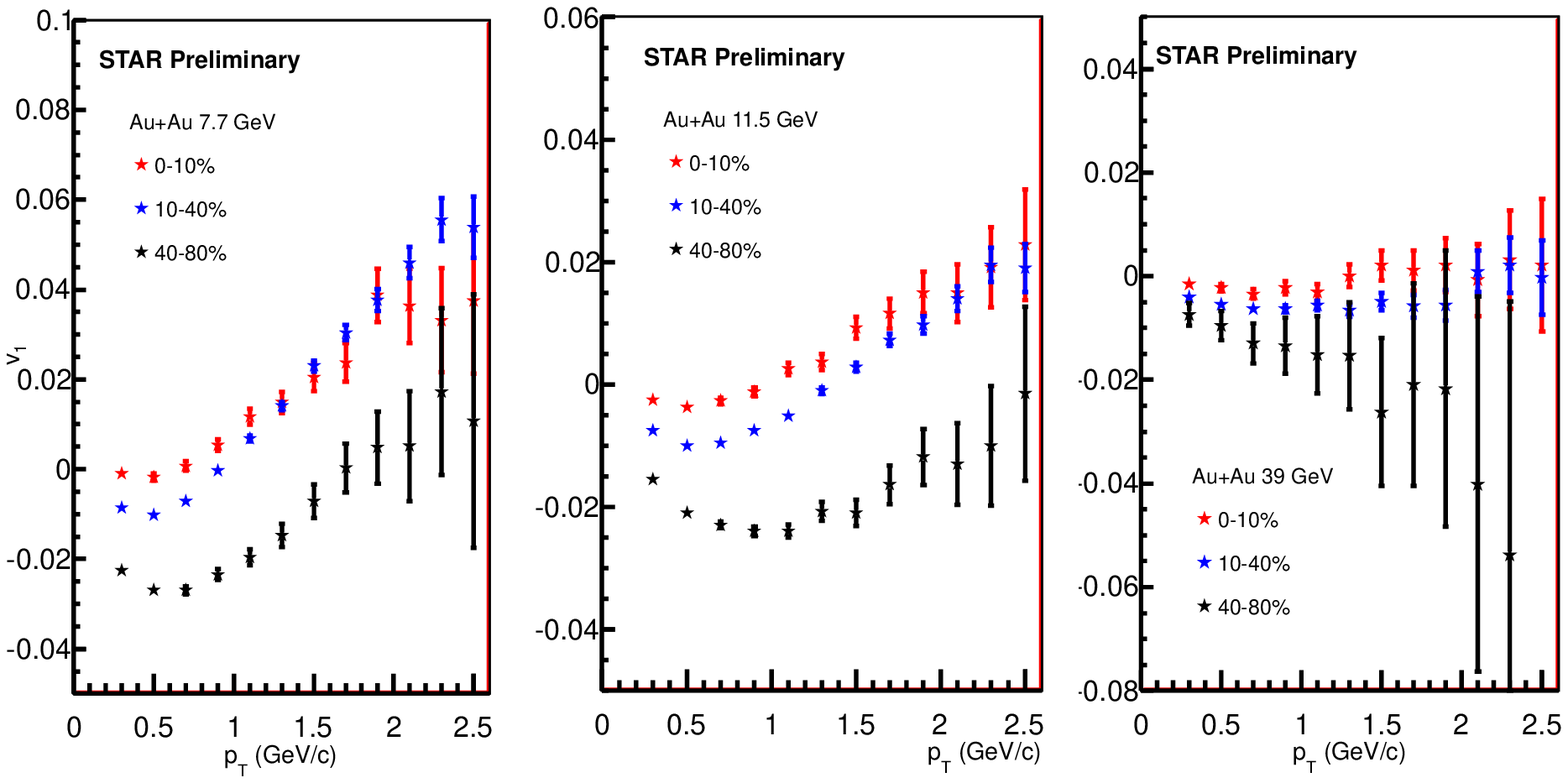}
\caption{Charged hadron $v_{1} $ as a function of transverse momentum  $p_{T} $ for central 0--10 \%, mid central 10--40\% and peripheral 40--80 \%  Au+Au collisions  at $\sqrt{s_{NN}}$  = 39, 11.5 and 7.7 GeV. The errors shown are statistical.  For this plot, particles from the main TPC are used with $ |\eta | < 1.0 $ }
\label{fig4}
\end{center}
\eef
Figure ~\ref{fig4} shows charged hadron $v_{1} $ as a function of transverse momentum  $p_{T} $ for central 0--10 \%, mid central 10--40\% and peripheral 40--80 \%  Au+Au collisions  at $\sqrt{s_{NN}}$  = 39, 11.5 and 7.7 GeV. The errors shown are statistical.  For this plot, particles from the main TPC are used with $ |\eta | < 1.0 $ . We observe the centrality dependence of directed flow as a function of $ p_{T} $ . We also observe that  the directed flow signal crosses zero at  $p_{T}  >1.0 GeV/c $ from negative to positive , which was previously reported  at higher RHIC energies \cite{v1-4systems} 

 \bef
\begin{center}
\includegraphics[scale=0.69]{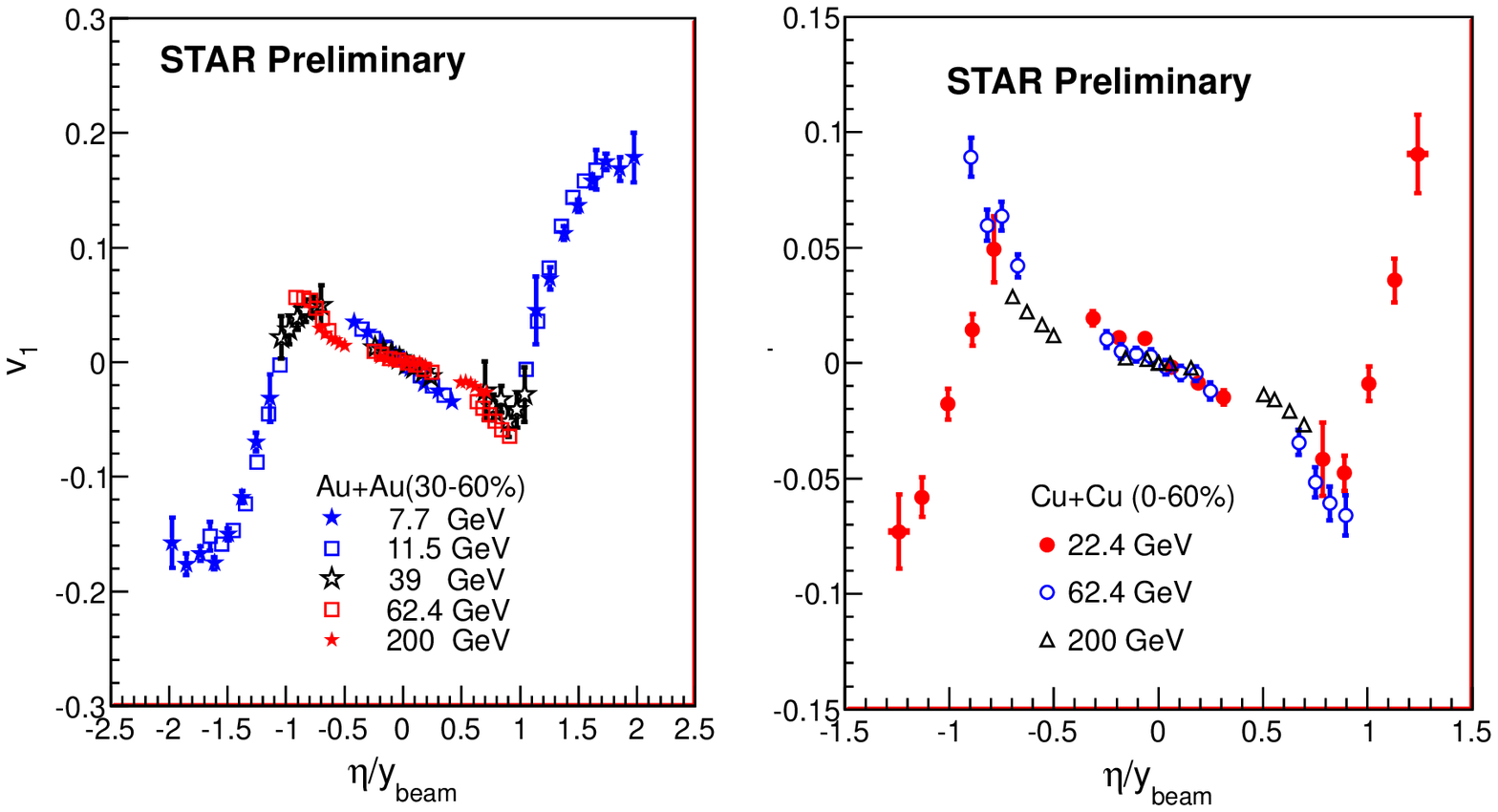}
\caption{ The left panel shows charged hadron $v_1$ as a function of $\eta$ scaled by the respective $y_{\rm{beam}}$  for Au+ Au Collisions at 7.7, 11.5, 39, 62.4 and 200 GeV  for 30--60 \% central collisions. The right panel shows charged hadron $ v_{1} $ as a function of $\eta$ scaled by the corresponding $y_{\rm{beam}}$ for Cu+Cu collisions at 22.4, 62.4 and 200 GeV.  The results for 62.4 and 200 GeV are for 30--60\% centrality, previously reported by STAR~\cite{v1-4systems}.}
\label{fig5}
\end{center}
\eef
Figure ~\ref{fig5} (left panel) shows charged hadron $v_1$ as a function of $\eta$ scaled by the corresponding $y_{\rm{beam}}$  for Au+Au Collisions at 7.7, 11.5, 39, 62.4 and 200 GeV  for 30--60 \% central collisions. The new results reported here are the charged hadron $v_1\{{\rm BBC}\}$ in Au+Au collisions for 30--60\% centrality at $\sqrt{s_{NN}} $ = 39,11.5 and 7.7 GeV.
On the right, we plot charged hadron $ v_{1} $ as a function of $\eta$ scaled by  the corresponding $y_{\rm{beam}}$ for Cu+Cu collisions at 22.4, 62.4 and 200 GeV.  The results for 62.4 and 200 GeV are  for 30--60\% centrality  previously reported by STAR~\cite{v1-4systems}.  The new results reported here are the charged hadron $v_1\{{\rm BBC}\}$ in Cu+Cu collisions for 0--60\% centrality at $\sqrt{s_{NN}} = 22.4$ GeV. We observe that  $v_1(\eta/y_{\rm beam})$  shows a beam energy scaling behavior, though not perfect, that has already been observed at higher RHIC energies. 

\bef
\begin{center}
\includegraphics[scale=0.69]{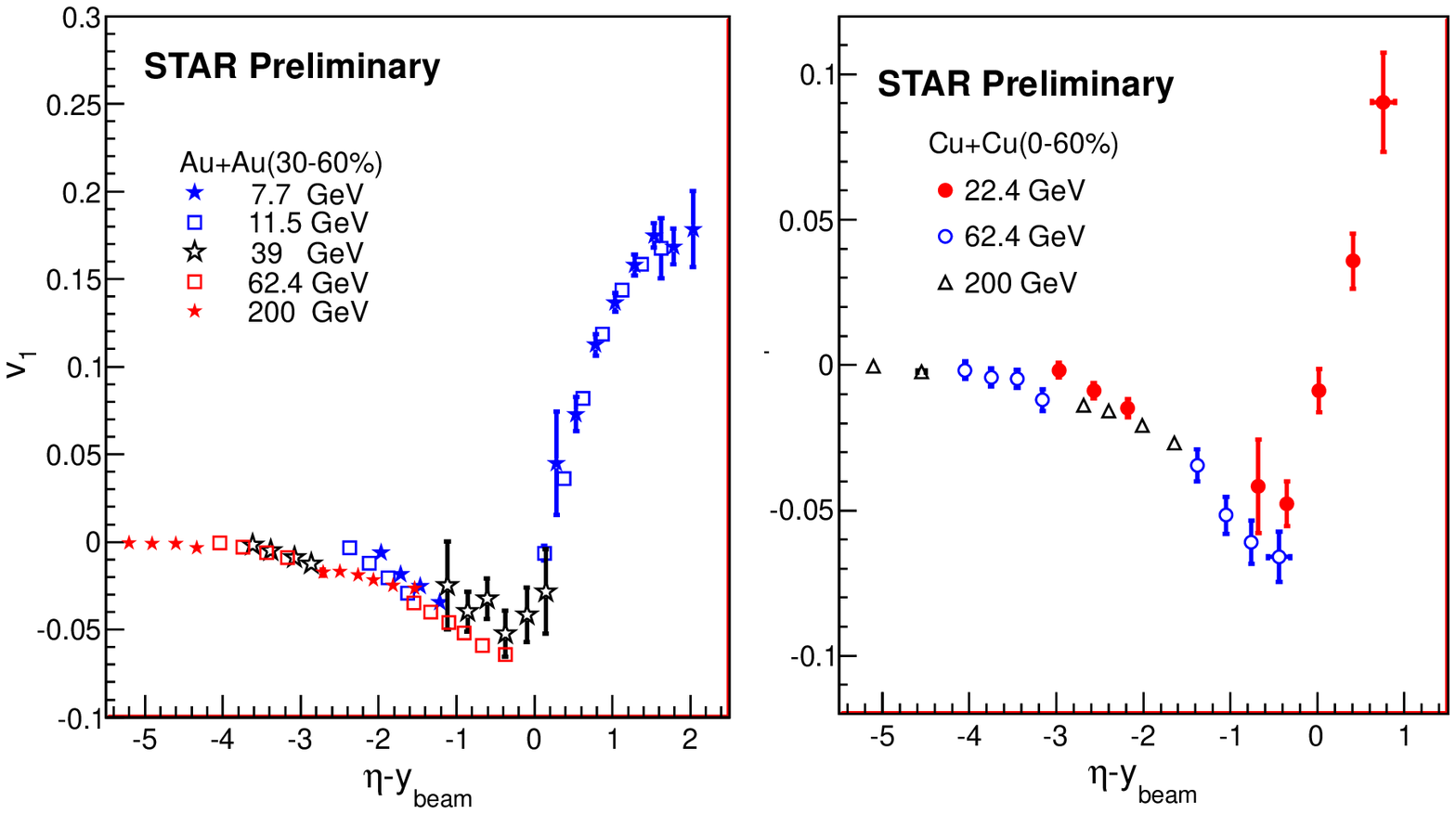}
\caption{ The left panel shows charged hadron $v_1$ as a function of $\eta - y_{\rm{beam}}$  for Au+Au collisions at 7.7, 11.5, 39, 62.4 and 200 GeV  for 30-60 \% centrality.  The right panel gives charged hadron $ v_{1} $ as a function of $\eta - y_{\rm{beam}}$ for Cu+Cu collisions at 22.4, 62.4 and 200 GeV.  The results for 62.4 and 200 GeV are for 30--60\% centrality, previously reported by STAR~\cite{v1-4systems}.}
\label{fig6}
\end{center}
\eef
Figure ~\ref{fig6} (left) shows  charged hadron $v_1$ as a function of $\eta$ - $y_{\rm{beam}}$  for Au+ Au Collisions at 7.7, 11.5, 39, 62.4 and 200 GeV  for 30--60 \% central collisions. On the right, we show charged hadron $ v_{1} $ as a function of $\eta - y_{\rm{beam}}$   for Cu+Cu collisions  22.4, 62.4 and 200 GeV.  The results for 62.4 and 200 GeV are  for 30--60\% centrality, previously reported by STAR~\cite{v1-4systems}. In this frame, zero on the horizontal axis corresponds to beam rapidity for each of the these beam energies. The data support the limiting fragmentation hypothesis~\cite{v1-4systems} in the region $-2.0 < y_{\rm{beam}}  < -1 $  for both system Au+ Au and Cu+ Cu presented here.

\subsection{Elliptic  flow}

\bef
\begin{center}
\includegraphics[scale=0.79]{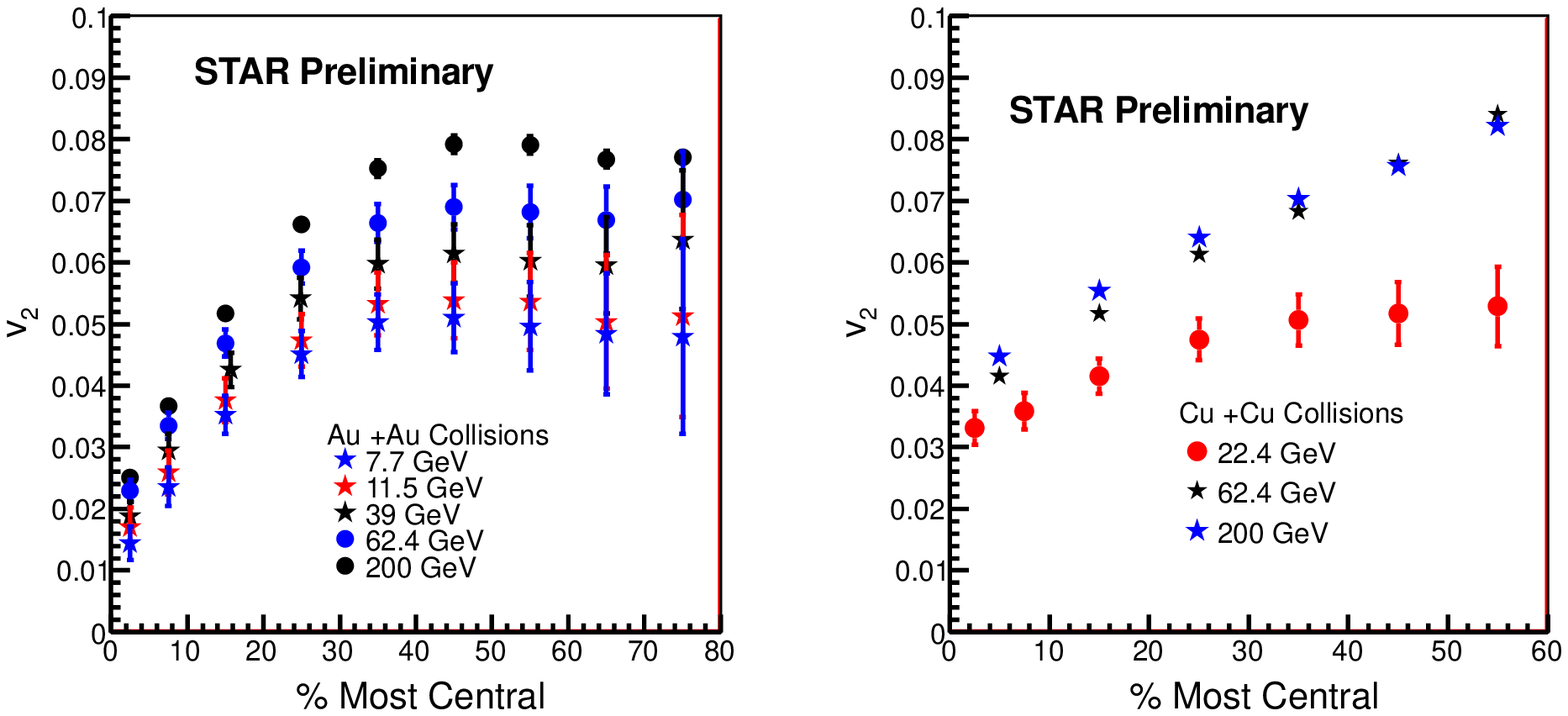}
\caption{The left panel shows elliptic flow $v_2$ as a function of centrality for charged hadrons from Au+Au collisions at $\sqrt{s_{NN}} $= 7.7, 11.5, 39, 62.4 and 200 GeV. The right panel shows elliptic flow $ v_{2} $ as a function of centrality for Cu+Cu collisions  at $\sqrt{s_{NN}} $=22.4, 62.4 and 200 GeV. These results are based on the two-particle direct cumulant method with $ |\eta |<1.0 $  and $0.2 <  p_{T} <  2.0 GeV $. }
\label{fig7}
\end{center}
\eef
Figure~\ref{fig7} (left) shows elliptic flow $v_2$ as a function of centrality for charged hadrons from Au+Au collisions at $\sqrt{s_{NN}} $= 7.7, 11.5, 39, 62.4 and 200 GeV. The right panel presents elliptic flow $ v_{2} $ as a function of centrality for Cu+Cu collisions  at $\sqrt{s_{NN}} $=22.4, 62.4 and 200 GeV. These results are based on the two-particle direct cumulant method with $ |\eta |<1.0 $  and $0.2 <  p_{T} <  2.0$ GeV.  We observe that $v_{2} $ as a function of centrality increases with beam energy.

\bef
\begin{center}
\includegraphics[scale=0.79]{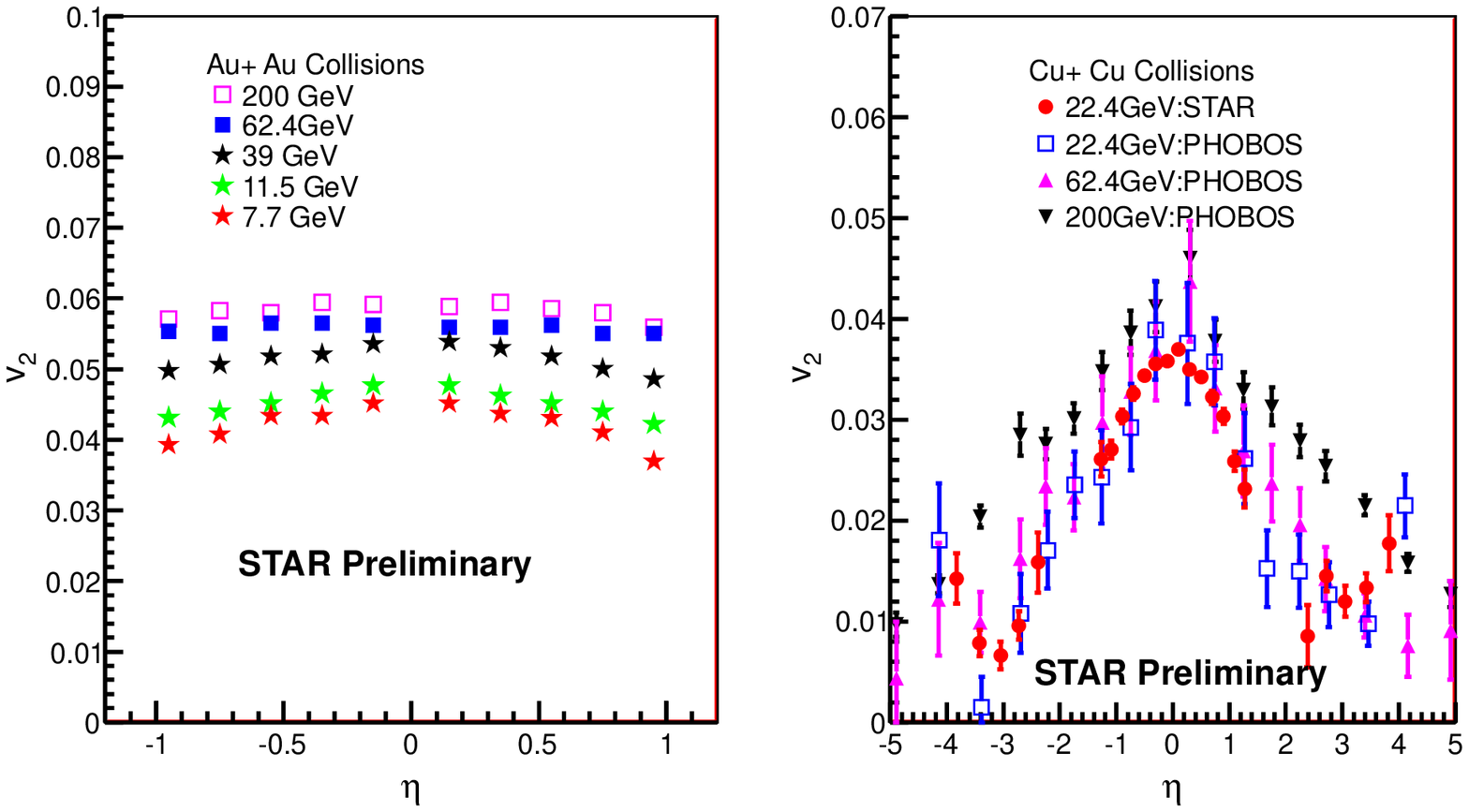}
\caption{The left panel shows $v_2$ as a function of $ \eta $  for Au+Au collisions at 7.7, 11.5, 39, 62.4 and 200 GeV for 0--40 \% central collisions.  On the right, we show the  elliptic flow versus $ \eta $ for 
charged hadrons from 22.4 GeV Cu+Cu collisions at 0--60\% centrality measured with the $\eta$-subevent method with an $\eta$ gap of 0.3 units.  Results are compared with data from the PHOBOS collaboration
 at  22.4, 62.4 and 200 GeV with 0--40 \%  centrality.} 
\label{fig8}
\end{center}
\eef

Figure~\ref{fig8}  left panel shows $v_2$ as a function of $ \eta $  for Au+Au collisions at 7.7, 11.5, 39, 62.4 and 200 GeV for 0--40 \% centrality.  On the right, we show elliptic flow versus $ \eta $ for charged hadrons from Cu+Cu collisions at 0--60\% centrality at $\sqrt{s_{NN}} = 22.4$ GeV compared with the results from the PHOBOS collaboration at  22.4, 62.4 and 200 GeV with 0--40 \%  centrality.  The new 22.4 GeV Cu+Cu results reported here are based on the $\eta$-subevent method with an $\eta$ gap of 0.3 units.  We observe that $v_{2}$  as a function of  $\eta$ increases with beam energy. This energy dependence of $p_{T}$  integrated elliptic flow as a function of centrality and $\eta$  is driven by the increase in mean $ p_{T } $ with energy.  

\bef
\begin{center}
\includegraphics[scale=0.69]{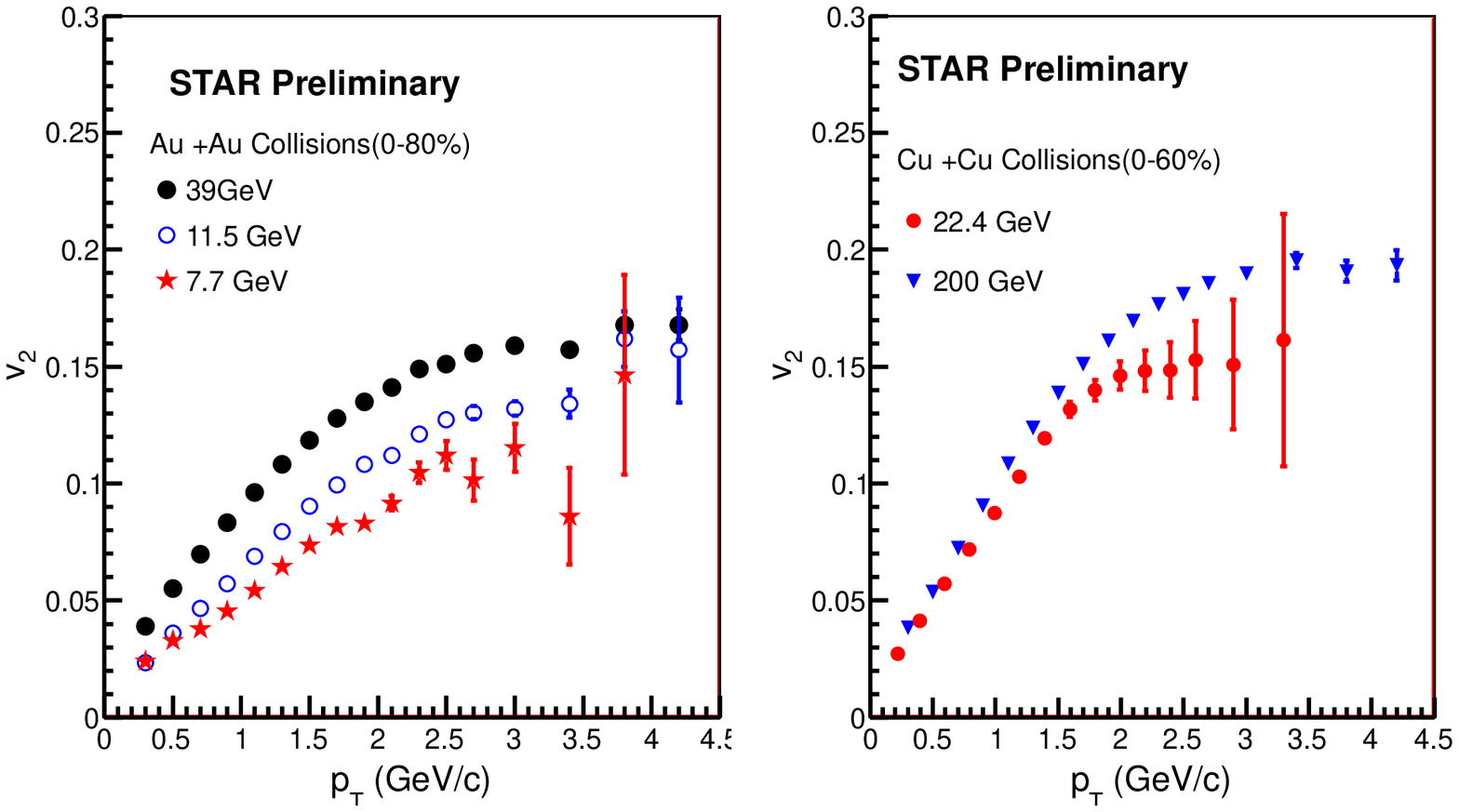}
\caption{The left panel shows $ v_{2}$ as a function of $ p_{T} $  for Au+Au collisions at 7.7, 11.5 and 39 GeV for 0--80 \% centrality. The right panel shows elliptic flow versus $ p_{T}  $ for charged hadrons from Cu+Cu collisions at 0--60\% centrality at $\sqrt{s_{NN}} = 22.4$ GeV measured with the $\eta$-subevent method with an $\eta$ gap of 0.3 units. Results are compared with previously  published data at 200 GeV.} 
\label{fig9}
\end{center}
\eef

Figure ~\ref{fig9} left panel shows $ v_{2}$ as a function of $ p_{T} $  for Au+Au collisions at 7.7, 11.5 and 39 GeV for 0-80 \% centrality. The results reported here are based on the event plane method using full event plane from TPC and are consistent with the result obtained with two particle cumulant methods. On the right, we show elliptic flow versus $ p_{T}  $ for charged hadrons from Cu+Cu collisions at 0--60\% centrality at $\sqrt{s_{NN}} = 22.4$ GeV, compared with the previously  published results at 200 GeV.  Again, the new 22.4 GeV Cu+Cu results reported here are based on the $\eta$-subevent method with an $\eta$ gap of 0.3 units.  The error bars include only statistical uncertainties. We observe $ v_{2} $ as a function of 
$p_{T}$   obtained using two particle correlation as well as event plane method for Au+Au collisions shows the energy dependence at lower beam energies upto  39 GeV. However the result obtained using four particle cumulant method \cite{fourptcorr} shows very small energy dependence. 
\bef
\begin{center}
\includegraphics[scale=0.59]{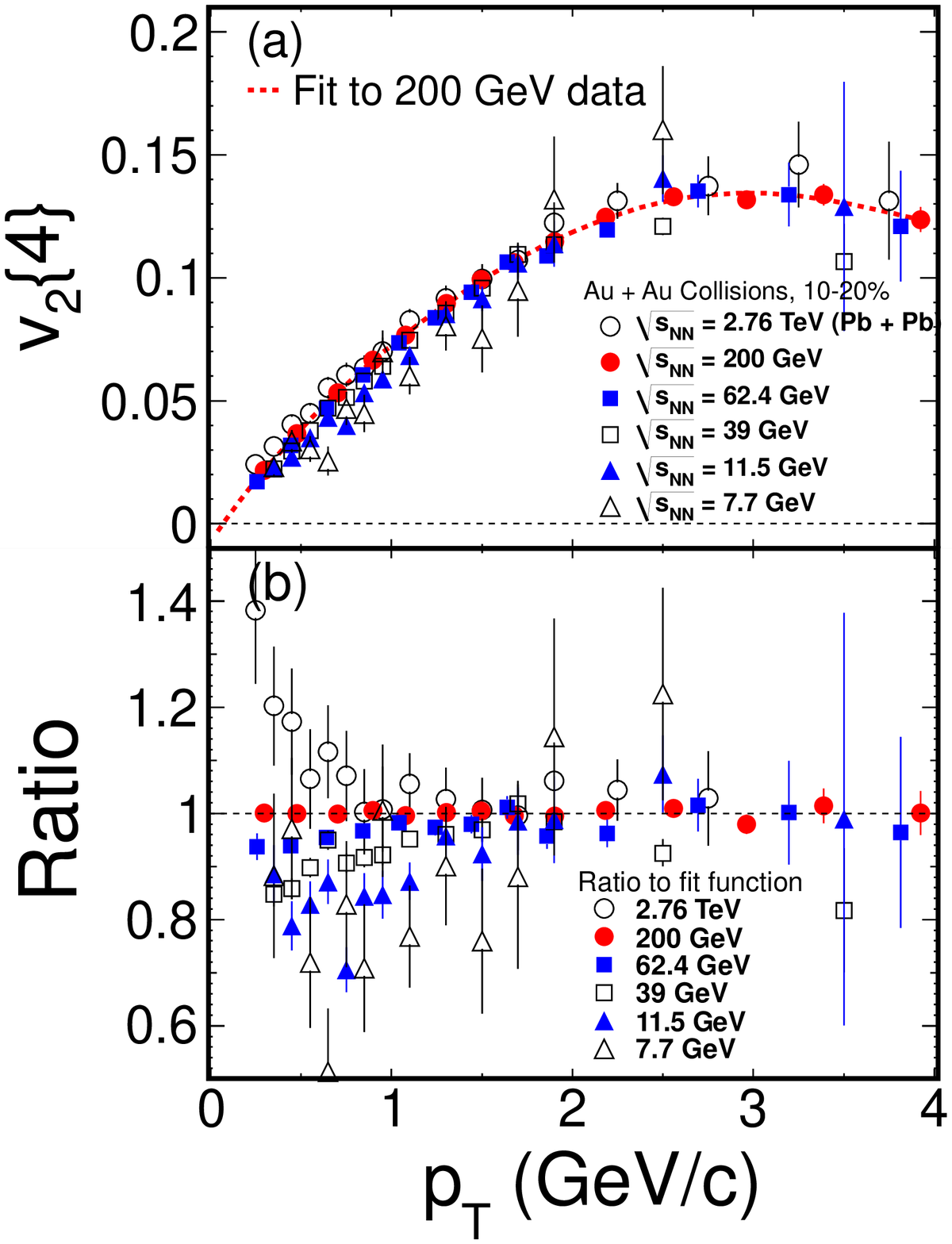}
\caption{Elliptic flow $v_2\{4\}$ for charged hadrons as a function of transverse momentum for various beam energies from 7.7 GeV to 2.76 TeV.
The results for 7.7 to 200 GeV are from Au+Au collisions and those for 2.76 TeV are from Pb+Pb collisions \cite{ALICE_Flow}. The red line shows a fit to the results  from Au+Au collisions at $\sqrt{s_{NN}} $= 200 GeV. The bottom panel shows $v_2\{4\}$ divided by the fitted line, plotted vs. $ p_{T} $  for all energies. The results are shown for 10--20 \% central collisions. }
\label{fig10}
\end{center}
\eef

Figure~\ref{fig10} presents elliptic flow $v_2\{4\}$ for charged hadrons as a function of transverse momentum for various beam energies from 7.7 GeV to 2.76 TeV. 
The results from 7.7 to 200 GeV are from Au+Au collisions at RHIC and those for 2.76 TeV are from Pb+Pb collisions at the LHC \cite{ ALICE_Flow}. The red line shows a fit to the results  from Au+Au collisions at $\sqrt{s_{NN}} $= 200 GeV. The bottom panel shows $v_2\{4\}$ divided by the fitted line, plotted vs. $ p_{T} $  for all energies. The results are shown for 10--20 \% central collisions.  It is observed that $v_2\{4\}(p_T)$ for all energies shows similar values 
beyond $p_T \sim 500$ MeV$/c$.  The agreement is within $\sim$ 10 \% of the 200 GeV data if we consider only energies of 39 GeV and above.  The statistical errors at the energies of 11.5 and 7.7 GeV are larger. This saturation is remarkable considering the very wide energy range involved.

\section{Summary}

In this  proceedings, we present STAR results for directed and elliptic flow as a function of transverse  momentum, pseudorapidity and centrality for Au+Au collisions at  39, 11.5 and 7.7 GeV 
and for Cu+Cu collisions at 22.4 GeV. Our findings extend to 22.4 GeV the observation that directed  flow is independent of system size (comparing Cu+Cu with Au+Au) at 62.4 and 200 GeV. 
Our findings also demonstrate that  $v_1(\eta/y_{\rm beam})$  shows a beam energy scaling behavior, though not perfect, that has already been established at higher RHIC energies. When 
viewed in the projectile frame of reference, our results are consistent with the limiting fragmentation hypothesis.  Elliptic flow as a function of centrality and pseudorapidity increases with beam 
energy in both systems studied, i.e.,  Au+Au and Cu+Cu. The centrality and  pseudorapidity  dependence are similar to that observed at higher RHIC beam energies.  The measured 
elliptic flow as a function of transverse momentum increases from 7.7 to 39 GeV but shows saturation above that energy up to 2.76 TeV. 

We thank the RHIC Operations Group and RCF at BNL, and the NERSC Center   at LBNL and the resources provided by the Open Science Grid consortium
for their support. This work was supported in part by the Offices of NP and HEP within the U.S. DOE Office of Science, the U.S. NSF, the Sloan 
Foundation, the DFG cluster of excellence ``Origin and Structure of the Universe", CNRS/IN2P3, RA, RPL, and EMN of France, STFC and EPSRC of the 
United Kingdom, FAPESP of Brazil, the Russian Ministry of Sci. and Tech., the NNSFC, CAS, MoST, and MoE of China, IRP and GA of the Czech Republic, 
FOM of the Netherlands, DAE, DST, and CSIR of the Government of India, the Polish State Committee for Scientific Research, and the Korea Sci. $ \& $  Eng. Foundation and Korea Research Foundation.


\section{References}

\begin{thebibliography}{99}

\bibitem{whitepapers}
   BRAHMS~Collaboration, I.~Arsene {\it et al.} 
   Nucl. Phys. A {\bf 757}, 1 (2005);
   PHOBOS~Collaboration, B. B.~Back {\it et al.} 
   Nucl. Phys. A {\bf 757}, 28 (2005);
   STAR~Collaboration, J.~Adams {\it et al.} 
   Nucl. Phys. A {\bf 757}, 102 (2005);
   PHENIX Collaboration, K.~Adcox {\it et al.} 
   Nucl. Phys. A {\bf 757}, 184 (2005).

\bibitem{collectiveflow}
         J.-Y. Ollitrault, Phys. Rev. D {\bf 46}, 229 (1992).

\bibitem{methods}
        A. M. Poskanzer  and S. A. Voloshin, Phys. Rev. C {\bf 58}, 1671 (1998).

\bibitem{Sorge}
   H. Sorge, Phys. Rev. Lett. {\bf 78}, 2309 (1997).

\bibitem{Herrmann}
   N.~Herrmann, J. P.~Wessels, and T.~Wienold,
   Annu. Rev. Nucl. Part. Sci. {\bf 49}, 581 (1999).

\bibitem{Hydro}
U. W. Heinz, arXiv:0901.4355.

\bibitem{Transport}  
S. A. Bass {\it et al.} Prog. Part. Nucl. Phys. {\bf 41}, 225 (1998).

\bibitem{bes} 
B. I. Abelev {\it et al.} (STAR Collaboration), STAR Note SN0493, 2009.

\bibitem{Csernai}
L. P. Csernai and D. Rohrich, Phys. Lett. B {\bf 458}, 454 (1999).

\bibitem{Brachmann}
J. Brachmann {\it et al.}, Phys. Rev. C {\bf 61}, 024909 (2000).

\bibitem{Stocker}
H. St\"ocker, Nucl. Phys. A {\bf 750}, 121 (2005).

\bibitem{Wiggle}
R. J. Snellings, H. Sorge, S. A. Voloshin, F. Q. Wang and N. Xu, 
Phys. Rev. Lett {\bf 84}, 2803 (2000).

\bibitem{starnim}
        K.~H.~Ackermann {\it et al.},
        Nucl. Instr. Meth. A {\bf 499}, 624 (2003).

\bibitem{BBC}
C. A. Whitten, Jr. STAR Collaboration, AIP Conf. Proc. {\bf 980}, 390 (2008).

\bibitem{trigger} 
               F.~S.~Bieser {\it et al.}, 
               Nucl. Instr. Meth. A {\bf 499}, 766 (2003)

\bibitem{startpc}
        M. Anderson {\it et al.},
        Nucl. Instr. Meth. A {\bf 499}, 659 (2003).

\bibitem{starftpc}
               K.~H.~Ackermann  {\it et al.},
               Nucl. Instr. Meth. A {\bf 499}, 713 (2003).

\bibitem{zdc_smd}
          B. Christie, S. White, P. Gorodetzky, D. Lazic, 
          STAR Note SN0175, 1994.  

\bibitem{momentum} 
   M. Anderson, P. M. Dinh, J. Y. Ollitrault, A. M. Poskanzer, S. A. Voloshin, 
   Phys. Rev. C {\bf 66}, 014901 (2002).

\bibitem{flow1} J. Adams {\it et al.} (STAR Collaboration),
                Phys. Rev. C {\bf 72}, 014904 (2005).

\bibitem{PHOBOS_v1} B. B. Back {\it et al.} (PHOBOS Collaboration), Phys. Rev. Lett. {\bf 97}, 012301 (2006).

\bibitem{v1-4systems}
B.~I.~Abelev {\it et al.} (STAR Collaboration),
Phys. Rev. Lett. {\bf 101}, 252301 (2008).

\bibitem{ampt}
   Z. W. Lin and C.-M. Ko,
   Phys. Rev. C {\bf 65}, 034904 (2002);
   L.-W. Chen, C.-M. Ko, J. Phys. G {\bf 31}, S49 (2005).

\bibitem{urqmd}
   S. A. Bass {\it et al.}, Prog. Part. Nucl. Phys. {\bf 41}, 225 (1998);
   M. Bleicher {\it et al.}, J. Phys. G {\bf 25}, 1859 (1999).

\bibitem{CuCuPaper}
B.~I.~Abelev {\it et al.} (STAR Collaboration),
Phys. Rev. C {\bf 81}, 044902 (2010).

\bibitem{vtwodiff}
 B.~I.~Abelev {\it et al.} (the STAR Collaboration),
Phys. Rev. C {\bf 75}, 054906 (2007).

\bibitem{PH0BOS_cu22} R. Nouicer for the PHOBOS Collaboration, J. Phys. G, Nucl. Part. Phys. {\bf 34}, S887 (2007).

\bibitem{PH0BOS_v2} B. Alver {\it et al.} (PHOBOS Collaboration), Phys. Rev. Lett. {\bf 98}, 242302 (2007).

\bibitem{fourptcorr}  Nicolas Borghini, Phuong Mai Dinh, and Jean-Yves Ollitrault ,Phys. Rev. C {\bf 63}, 054906 (2001).

\bibitem {ALICE_Flow} K. Aamodt et al. (ALICE Collaboration), arXiv:1011.3914 [nucl-ex].
\end{thebibliography}
\end{document}